\documentstyle[12pt]{article}
\begin{document}

\bibliographystyle{unsrt}
\def\Journal#1#2#3#4{{#1} {\bf #2}, #3 (#4)} 
\def\NCA{\em Nuovo Cimento} 
\def\NIM{\em Nucl. Instrum. Methods} 
\def\NIMA{{\em Nucl.Instrum. Methods} A} 
\def\NPB{{\em Nucl. Phys.} B}
\def\PLB{{\em Phys. Lett.}B} 
\def\PRL{\em Phys. Rev. Lett.}
\def\PRD{{\em Phys. Rev.} D}
\def\ZPC{{\em Z. Phys.} C}
\def\MPLA{{\em Mod. Phys.Lett} A}
\def\ANP{\em Ann. Phys.} 
\def\PTP{\em Prog. Theor. Phys.} 
\def\PR{\em Phys. Rep.} 
\def\st{\scriptstyle}
\def\sst{\scriptscriptstyle}
\def\mco{\multicolumn}
\def\epp{\epsilon^{\prime}}
\def\vep{\varepsilon}
\def\ra{\rightarrow}
\def\ppg{\pi^+\pi^-\gamma}
\def\vp{{\bf p}}\def\ko{K^0}
\def\kb{\bar{K^0}}
\def\al{\alpha}
\def\ab{\bar{\alpha}}
\def\be{\begin{equation}}
\def\ee{\end{equation}}
\def\bea{\begin{eqnarray}}
\def\eea{\end{eqnarray}}
\def\CPbar{\hbox{{\rmCP}
\hskip-1.80em{/}}}

\noindent\hbox to\textwidth{February 1998 \hfill BROWN-HET-1110}\\
\noindent\hbox to\textwidth{ \hfill BROWN-TA-555}\\
\noindent\hbox to\textwidth{ \hfill KIAS-P98004}\\
\noindent\hbox to\textwidth{ \hfill hep-ph/9802xxx}\\
\vskip 0.6in
\begin{center}
{\bf Neutrino Mass Matrix, Mixing and Oscillation}\\
\vskip 0.2in
 Kyungsik Kang\footnote{supported in part by the 
USDOE contract DE-FG02-91ER 40688-Task A.} \\ {\small 
Department of Physics, Brown University,
Providence, RI 02912, USA}\\
and\\
Sin Kyu Kang \\ {\small
School of Physics, Korea Institute for Advanced Study, Seoul, Korea}\\
\vskip 0.2in
%
%

{\bf Abstracts}
\end{center}
\begin{quote}
We investigate the phenomenological consequences of lepton mass
matrices originated from the family permutation symmetry and its 
suitable breakings. Adopting the recently proposed new  mass matrix 
for the charged lepton mass matrix and the Majorana neutrino mass 
matrix considered before, we find that the resulting lepton flavor
mixing  matrix  is consistent with the current data on various type
of  neutrino oscillation experiments except the Chlorine data and
LSND measurement and that three neutrinos are almost degenerate
in mass. Numerically the three neutrinos can account for about $50\%$ 
of the hot dark matter(HDM) when the neutrino mass matrix is 
constrained by the small angle MSW solution for the solar neutrino 
problem, atmospheric neutrino data of the Super-Kamiokande group, 
and the limitation on the electron-type neutrino mass due to 
non-observation of neutrino-less double beta decays.
\end{quote}
\begin{center}
 (Presented by Kyungsik Kang\footnote{Presented at the APCTP Workshop: {\it Pacific Particle Physics Phenomenology}, Seoul, Korea, 31 October - 2 November, 1997})
\end{center}

\section{Introduction}
The flavor problem concerning the fermion masses, their hierarchical patterns,
as well as the flavor mixing
remains to be one of the most fundamental problems in particle 
physics.  Various attempts toward the understanding of the flavor mixing
with the quark mass matrix ansatz satisfying the calculability
condition are presented in another session \cite{skk}
Though the simple Fritzsch-type of mass matrix had attracted a great
deal of attention, it has been ruled out\cite{kangt} because
it  predicts the top quark mass to be below 
100 GeV.

Nevertheless, the Fritzsch-type mass matrix is very attractive
due to its simplicity.
Thus, the next move  is to generalize and modify the Fritzsch-type mass matrix
by maintaining the
calculability  property.   
Recently, we have proposed a generalized mass matrix ansatz and made
a systematic phenomenological study \cite{kang2}.
The general mass matrix ansatz leads to a hermitian mass matrix
\begin{eqnarray}
   M_H = \left( \begin{array}{ccc}
               0 & A & 0 \\
               A & D & B \\
               0 & B & C  \end{array} \right)
\end{eqnarray}
which is consistent with experiments for the range of $|w=B/D|=0.97 ~-~ 1.87$.
The case of $D=0$ reduces to the original Fritzsch type.
As shown in Refs. 1 and 3,
this form can be achieved by successive breaking of the ''democratic
mass matrix" that has the
maximal permutation symmetry, and is the generalization of
various specific forms of mass matrices proposed by others 
as special cases where $D$ and $B$ are related to each other 
in a particular way but mostly outside the range of $w$ that we found.
We  have shown that the matrix (1) with the finite range of 
$w$ and CP phase is consistent with experimental results including 
heavy top quark mass and the maximal CP violation.
%
We will assume the same form of mass matrices 
for the charged leptons sector.

On the other hand, all neutrino masses are zero and lepton numbers
are exactly conserved in the context of the standard model(SM).
Strictly speaking, it has not been established yet that the
neutrino masses are non-zero and hierarchical.  
However, the current experimental anomalies of 
solar \cite{homestake,gallex,sage,kamioka}
and atmospheric \cite{kamioka2,soudan,imb} neutrinos lead
us to speculate that neutrinos may be massive and have mixing,
as they can be interpreted as neutrino oscillations.
If we assume that only two neutrino flavors 
participate in the oscillations,
the parameter space consists of one mixing angle and one
mass-squared difference.
On the other hand, oscillations among three-neutrino flavors 
may provide a simultaneous solution for solar and atmospheric
neutrino problems.
In this case, we have two degrees of freedom in the
choice of mass-squared difference.
Several authors have suggested \cite{kkkk,degenerate}
that almost degenerate neutrinos are needed to accommodate
the solar and atmospheric neutrino observations as well as
the cosmological constraint
that arises when we regard neutrinos as  candidates for
the hot dark matter (HDM) within the three-flavor framework.

We will show that almost degenerate scenario
among three flavors indeed follows from the standpoint
of mass matrix ansatz.
In order to do so, we will 
construct neutrino mass matrices that are constrained by
the solar and atmospheric neutrino deficits and 
the non-observation of the neutrinoless double beta decay.
We note that both charged 
and neutral lepton mass matrices assumed above can originate from the 
permutation symmetry and its suitable breakings.

\section{Mass matrix for charged lepton}
Let us start with the new class of mass matrix Eq.(1) which is obtained from
the ``democratic mass matrix" with the maximal
$S(3)_{L}\times S(3)_{R}$ symmetry which is broken successively down to
$S(2)_L\times S(2)_R$ and $S(1)_L \times S(1)_R$, followed by a
unitary transformation with  
$ U=(u_1^{T}, u_2^{T}, u_3^{T})$, where $u_1=(\frac{1}{\sqrt{2}},
\frac{1}{\sqrt{6}}, \frac{1}{\sqrt{3}}), 
u_2=(-\frac{1}{\sqrt{2}},  \frac{1}{\sqrt{6}}, 
\frac{1}{\sqrt{3}}), u_3=(0, -\frac{2}{\sqrt{6}}, \frac{1}{\sqrt{3}} )$.
As discussed in Refs. 1 and 3, the (2,2) element is related to (2,3) element
in Eq. (1) by $w\equiv B/D = (k+1)/\sqrt{2}(2k-1) $ in the
hierarchical mass eigenstate.

Because of the quark-charged lepton symmetry, 
the mass matrix for the charged lepton sector is assumed to be exactly the same as
the quark mass matrix (1).
The mass matrix $M_H$ for $K=U^l_LU^{l\dagger}_R=diag[-1,1,1]$ 
can then be written as \cite{skk,kang2}
\begin{eqnarray}
   M_H = \left( \begin{array}{ccc}
               0 & \sqrt{\frac{m_1 m_2 m_3}{m_3-\epsilon}} & 0 \\
               \sqrt{\frac{m_1 m_2 m_3}{m_3-\epsilon}} & 
               m_2-m_1+\epsilon & w(m_2-m_1+\epsilon) \\
               0 & w(m_2-m_1+\epsilon ) &
               m_3-\epsilon  \end{array} \right),
\end{eqnarray}
in which the small parameter $\epsilon$ is related to $w$, i.e.,
$w \simeq \pm \frac{\sqrt{\epsilon m_3}}{m_2}
\left(1+\frac{m_1}{m_2}-\frac{m_2}{2m_3}\right)$,
whose range is to be determined from the experiments.

Then, the diagonalizing matrix $U_L^{l}$ can be written as
\begin{equation}
U_L^{l} = U_{23}(\theta_{23}) \cdot U_{12}(\theta_{12})
\end{equation}
where
\begin{eqnarray}
U_{12} =\left( \begin{array}{ccc}
              \cos \theta_{12} & \sin \theta_{12} & 0 \\
              -\sin \theta_{12} & \cos \theta_{12} & 0 \\
              0 & 0 & 1 \end{array} \right),
 \qquad
U_{23} =\left( \begin{array}{ccc}
              1 & 0 & 0 \\
              0 & \cos \theta_{23} & \sin \theta_{23}  \\
              0 & -\sin \theta_{23} & \cos \theta_{23} \end{array} \right)
\end{eqnarray}
Since (1,1), (1,3) and (3,1) elements of $M_H$ are zero, 
we may put $U_{13}(\theta_{13})=1$ 
without loss of generality.
The mixing angles $\theta_{12}$ and $\theta_{23}$ can be
written to a very good approximation as
$\tan \theta_{12}=\sqrt{\frac{m_1}{m_2}}$
and
\begin{equation}
\tan \theta_{23}=\frac{1}{2w}\left[\left(1+\frac{m_1-m_2}{m_3}\right)
-\sqrt{\left(1+\frac{m_1-m_2}{m_3}\right)^2+4w^2\left(\frac{m_1-m_2}
 {m_3}\right)}\right]
\end{equation}

For a negative $D$, 
the real symmetric  matrix $M_H$ can be diagonalized as 
$U_L^{l}M_H U_L^{l^{\dagger}}
= diag[m_{1}, -m_{2}, m_{3}]$, 
thus reversing the signs of both $m_1$ and $m_{2}$ in the above equations.

In both cases above
it turns out that the experimentally allowed range of $w$ for the quark
sector is $0.97 \leq  |w| \leq 1.87 $ in the leading
approximation \cite{kang2}.  
Thus, the same range of $w$ is assumed also for the charged lepton sector.
However, physical observables such as survival and transition 
probabilities for $\nu_{\alpha}$'s turn out to be
insensitive to the precise value of 
$w$ in the allowed range, as discussed 
in the following.  

\section{Mass matrix for neutrinos}
It is evident that the mass matrix of the charged lepton sector is not 
appropriate for the neutrino sector.
The neutrino oscillation experiments do not seem to support such 
hierarchical pattern for neutrino masses unlike the quark and charged 
lepton masses 
but rather nearly degenerate neutrinos
within the three-flavor framework \cite{kkkk,degenerate}.
In this section, we will show how such an almost degenerate neutrino scenario
follows from the neutrino mass matrix,
which is clearly different from the approach
used by others \cite{degenerate}.
We proposed for the neutrino sector the permutation symmetry among
three family indices rather than the flavor democracy.
It can be represented by the following matrix on the {\it symmetry basis}
of $S(3)$\cite{kkkk},
\begin{eqnarray}
   {\widetilde M}_{\nu}^{(0)} = c_{\nu}\left( \begin{array}{ccc}
               1 & r & r \\
               r & 1 & r \\
               r & r & 1  \end{array} \right).
\end{eqnarray}
After the unitary transformation with $V=(v^T_1, v^T_2, v^T_3)$ where
$v_1=(\frac{1}{\sqrt{3}}, \frac{1}{\sqrt{6}}, \frac{1}{\sqrt{2}})$,
$v_2=(\frac{1}{\sqrt{3}}, \frac{1}{\sqrt{6}}, -\frac{1}{\sqrt{2}})$
and
$v_3=(\frac{1}{\sqrt{3}}, -\frac{2}{\sqrt{6}}, 0),$
 the matrix becomes (6)
\begin{eqnarray}
    M_{\nu}^{(0)} = c_{\nu}\left( \begin{array}{ccc}
               1+2r & 0 & 0 \\
               0 & 1-r & 0 \\
               0 & 0 & 1-r  \end{array} \right).
\end{eqnarray}
Note that all three neutrinos are degenerate either for $r=0$ or
$r=-2$, while two neutrinos are always degenerate for any $r$ \cite{kkk}, 
so that only one $\Delta m^2_{\nu} $ is available.

In order to confront the above mass matrix ansatz with the experimental
and cosmological observations, we make three observations for neutrinos 
which may be accounted for by assuming massive neutrinos: \\
(1) solar neutrino data from four different experiments,
the HOMESTAKE \cite{homestake}, GALLEX \cite{gallex}, SAGE \cite{sage},
and the KAMIOKANDE II-III \cite{kamioka}, (2)
atmospheric neutrino data measured by three experiments,
the Super-KAMIOKANDE \cite{kamioka2}, SOUDAN2 \cite{soudan}, and IMB \cite{imb},
(3) the likely need of neutrinos as a candidate of dark matter \cite{dark}.
As is well known, the solar neutrino deficit can be explained through
the MSW mechanism if $\Delta m_{solar}^{2} \simeq
6\times 10^{-6}~{\rm eV}^2$ and $\sin^{2} 2\theta_{solar} \simeq 7 \times 10^{-3}$
(small angle case) , or $\Delta m_{solar}^2 \simeq 9\times 10^{-6}~{\rm eV}^2$
and $\sin^{2} 2\theta_{solar} \simeq 0.6$ (large angle case)
or through the just-so vacuum oscillations if $\Delta m_{solar}^{2} \simeq
10^{-10}~{\rm eV}^2$ and $\sin^{2} 2\theta_{solar} \simeq 0.9$ \cite{PDG}.
The atmospheric neutrino  data of the Super-Kamiokande is consistent with
$\Delta m_{32}^2 \approx (2-8)\times 10^{-3} eV^2$ and 
$\sin^2\theta_{32}=0.85-1,$ at $90\%$CL.
If the light neutrinos account for the HDM of
the universe, one has to require \cite{dark}
$\sum_{i=1,2,3} | m_{\nu_i} | \sim 6~{\rm eV}.$
Thus we see that all three neutrinos may be almost 
degenerate in their masses, with $m_{\nu_i} \sim $~ a few eV, rather than
$m_{\nu_1} \ll m_{\nu_2} \ll m_{\nu_3}$, as assumed sometimes in the 
three-neutrino mixing scenarios.
As discussed above, in order to account for the solar and atmospheric
neutrino oscillations simultaneously, two separate $\Delta m^2_{ij}
=m^2_{\nu_i}-m^2_{\nu_j}$ 
scales are required.  
The two $\Delta m^2_{ij}$ scales can be generated from our neutrino mass 
matrix (7) in such a way to lift the degeneracy 
between the second and third neutrinos, i.e., by breaking $S(3)$ to $S(2)$.
It can be achieved by allowing (3,2) and (2,3) elements of matrix (7)
to be nonzero.
Then, the resulting matrix for the neutrinos in the flavor basis becomes
\begin{eqnarray}
    M_{\nu} = c_{\nu}\left( \begin{array}{ccc}
               1+2r & 0 & 0 \\
               0 & 1-r & \epsilon_{\nu} \\
               0 & \epsilon_{\nu} & 1-r  \end{array} \right),
\end{eqnarray}
which can be diagonalized by 
$ U_{\nu}=(u_1^{T}, u_2^{T}, u_3^{T})$, where 
$u_1=(1, 0, 0 ),
u_2=(0, \frac{1}{\sqrt{2}}, \frac{1}{\sqrt{2}}), 
u_3=(0,\frac{1}{\sqrt{2}},  -\frac{1}{\sqrt{2}})$,
with the resulting eigenvalues
$m_{\nu_{i}} / c_{\nu} = (1+2r), (1-r \mp \epsilon_{\nu}).$
Note that $U_{\nu}$ is independent of neutrino mass 
eigenvalues, or equivalently, of $c_{\nu}, \epsilon_{\nu}$ and $r$. 
This is because the mass matrix $M_{\nu}$ given by (8) 
has a residual
$S(2)$ symmetry acting upon the first and the second family indices.  
Alternatively, one may take nonzero (2,1) and (1,2) elements of the
matrix (7), but such case turns out to be inconsistent with the
atmospheric data from Super-Kamiokande.

One can solve for $c_{\nu}, r$ and $\epsilon_{\nu}$
by requiring three conditions, $\Delta m_{solar}^2 = 10^{-5}~
{\rm eV}^2$, $\Delta m_{atmos}^2 = 10^{-2} ~{\rm eV}^2$ 
and the neutrino mass bound due to non-observation of the neutrino-less
double beta decay\cite{doublebeta}, $<m_{\nu}>=|\sum^{3}_{i=1}
\eta_iV^2_{\nu_1i}m_i| \leq 1\sim 2$ eV, where $\eta_i=\pm1$ depending
on the CP property of $\nu_i$.
Note that we solve the solar
neutrino problem by adopting the MSW mechanism.
Then we get $r\simeq -1.9925, c_{\nu} \simeq 0.3345$ and 
$\epsilon_{\nu} \simeq 0.0075$ leading
to $|m_1|\simeq 0.998301 ~\mbox{eV}, |m_2| \simeq 0.998305 ~\mbox{eV}$ and 
$|m_3|\simeq 1.0033 ~\mbox{eV}$.
We note that the breaking parameter $\epsilon_{\nu}$
turns out to be much smaller than the parameter $r$

\section{Neutrino mixing matrix and predictions}
Combining  the $U^{l}_L$  given by Eq.(3) with $U_{\nu}$ diagonalize the
neutrino mass matrix (8), 
we get the flavor mixing matrix in the lepton sector,
\begin{eqnarray}
V_{\nu} \equiv U_{L}^{l^{\dagger}} ~U_{\nu}
=\frac{1}{\sqrt{2}}\left( \begin{array}{ccc}
               \sqrt{2}c_{12} &
                s_{12}(-c_{23}+s_{23} &
                s_{12}(c_{23}+s_{23}) \\
               \sqrt{2}s_{12} &
                c_{12}(c_{23}-s_{23}) &
                c_{12}(-c_{23}-s_{23}) \\
               0 & s_{23}+c_{23} & -s_{23}+c_{23}
                \end{array} \right),
\end{eqnarray}
where we have abbreviated $\cos\theta_{ij}$ and $\sin\theta_{ij}$ as
$c_{ij}$ and $s_{ij}$ respectively.
For the whole range of $0.97 \leq |w| \leq 1.87 $, 
the neutrino mixing matrix is given by
\begin{eqnarray}
    |V_{\nu}| = \left( \begin{array}{ccc}
               0.9952 & -0.0453 & 0.0521  \\
               0.0692 & 0.6521 & -0.7494 \\
               0.0000 & 0.7531 & 0.6551
          \end{array} \right) \leftarrow 
 \qquad
      \rightarrow
     \left( \begin{array}{ccc}
               0.9952 & -0.0440 & 0.0506  \\
               0.0692 & 0.6326 & -0.7271 \\
               0.0000 & 0.7307 & 0.6356
          \end{array} \right).
\end{eqnarray}
Note that our lepton mixing matrix predicts zero for $(V_{\nu})_{31}$ 
element, i.e., the tau-$\nu_{e}$ coupling is forbidden.
Because of the smallness of the (1,2) and (1,3) elements in (10),
$<m_{\nu}> \simeq m_1 \simeq 1$ eV.
>From this mixing matrix, we obtain $\sin^2{2\theta_{12}} \simeq 0.0054 ~-~
0.0061$ which is consistent with the small angle MSW solution
 and $\sin^2{2\theta_{23}} \simeq 0.85 ~-~ 0.96$ consistent with the 
Super-Kamiokande result.

The predicted probability of $\nu_{e} ~-~ \nu_{\mu}$ oscillation
is given by $P(\nu_{e}-\nu_{\mu})\sim 4|V_{e3}|^2|V_{\mu 3}|^2
sin^2\left(\frac{1.27\Delta m^2_{32}L}{E}\right) \sim 
0.006~sin^2\left(\frac{1.27\Delta m^2_{32}L}{E}\right)$.
In addition, we see that three neutrino can account for about
$50\%$ of the HDM becaus we get $\sum|m_{\nu_i}| \sim 3 $ eV whereas
HDM mass limit is 6 eV.

\section{Conclusion}
In conclusion, we investigated in this paper phenomenological consequences of 
the lepton mass matrix ansatzs with the minimal 
number of parameters, three each in the charged lepton 
and Majorana neutrino mass matrices $M_{H}$ and $M_{\nu}$, 
with permutation symmetries among three 
generations $S(3)_L \times S(3)_R$ and $S(3)$ that are broken down to
$S(2)_L \times S(2)_R$ and $S(1)_L \times S(1)_R$ and $S(2)$ and $S(1)$, 
respectively.
We find the mass matrix ansatz Eqs.(1) and (8) lead to a lepton mixing matrix which can
accommodate the solar and atmospheric neutrino observations as well as
the constraint from
non-observation of neutrino-less beta decays.
Three light Majorana neutrinos can  provide about $50\%$ of the 
HDM, with $\Sigma | m_{\nu_i} | \sim 6~$ eV. 

{\it Note added }: After the presentation of this talk,  
the results of the long-baseline
reactor experiment CHOOZ appeared \cite{chooz}, which are consistent with
our results.
\end{document}